\documentclass{article}

\usepackage{longcat_codec}

\usepackage[utf8]{inputenc} % allow utf-8 input
\usepackage[T1]{fontenc}    % use 8-bit T1 fonts
\usepackage{hyperref}       % hyperlinks
\usepackage{url}            % simple URL typesetting
\usepackage{booktabs}       % professional-quality tables
\usepackage{amsfonts}       % blackboard math symbols
\usepackage{amsmath}        % for \textnormal and other math formatting
\usepackage{nicefrac}       % compact symbols for 1/2, etc.
\usepackage{microtype}      % microtypography
\usepackage{xcolor}         % colors
\usepackage{graphicx}       % pictures
\usepackage{multirow}
\usepackage{array}
\usepackage{enumerate}
\usepackage{threeparttable} % 用于添加表格注释
\usepackage{adjustbox}
\usepackage{makecell}
\usepackage{appendix}  % 用于管理附录格式
\usepackage{natbib} %引用格式上，考虑到table5和6在密集引用文献，格式还是保留[x]这种简单的
\usepackage{caption}  % 加载 caption 宏包精细控制
\title{LongCat-Audio-Codec: An Audio Tokenizer and Detokenizer Solution Designed for Speech Large Language Models}

\author{%
\normalfont Xiaohan Zhao, Hongyu Xiang, Shengze Ye, Song Li,\\
\normalfont Zhengkun Tian, Guanyu Chen\thanks{work done during internship at Meituan}, Ke Ding, Guanglu Wan \\
\\
\textbf{LongCat Team, Meituan}
} 

\renewcommand{\headeright}{\raisebox{-0.2\height}{\includegraphics[height=1.8em]{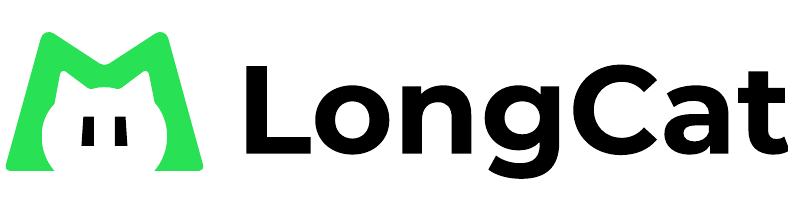}}}
\renewcommand{\shorttitle}{LongCat-Audio-Codec}

\begin{document}

\maketitle

\begin{abstract}
  This paper presents LongCat-Audio-Codec, an audio tokenizer and detokenizer solution designed for industrial grade end-to-end speech large language models. 
  By leveraging a decoupled model architecture and a multistage training strategy, LongCat-Audio-Codec exhibits robust semantic modeling capabilities, flexible acoustic feature extraction capabilities, and low-latency streaming synthesis capabilities. It encodes speech at an ultra-low frame rate of 16.67 Hz, with a minimum bitrate of 0.43 kbps and a maximum bitrate of 0.87 kbps. Evaluation results demonstrate that LongCat-Audio-Codec achieves strong speech intelligibility and is capable of synthesizing high-quality speech at low bitrate, thus effectively balancing coding efficiency and decoding quality.
  The inference code and model checkpoints of LongCat-Audio-Codec are available at: {\url{https://github.com/meituan-longcat/LongCat-Audio-Codec}}.
\end{abstract}

\section{Introduction}
\label{intr}
With the advancement of Speech LLM, the technical paradigm has gradually evolved to be grounded in discrete speech tokens.
An effective speech tokenizer must balance the needs of both understanding and generation tasks, incorporating sufficient information while maintaining a compact representation to facilitate model learning.
Speech tokens are divided into two main categories: acoustic tokens and semantic tokens, distinguished by the type of information they encode. 
Acoustic tokens are generated by neural codecs such as Encodec\,\citep{defossez2022encodec} and DAC\,\citep{kumar2024dac}. 
These tokens capture fine-grained acoustic characteristics of raw waveforms, offering an alternative to conventional frame-level features such as mel-spectrograms. 
On the other hand, semantic tokens are obtained by quantizing speech features rich in contextualized linguistic information, sourced from self-supervised speech models\,\citep{borsos2023audiolm,lakhotia2021gslm,nguyen2024spirit} or speech recognition systems. 
Besides, supervised semantic tokenization\,\citep{du2024cosyvoice,du2024cosyvoice2,du2025cosyvoice3} has been proposed, where semantic tokens are extracted from an ASR system or audio reasoning system by integrating vector quantization or scalar quantization into the encoder.

The development of acoustic codecs primarily encompasses the following directions:
Model scale-up, as exemplified by StableCodec\,\citep{Julian2024StableCodec} and WavTokenizer\,\citep{Ji2024WavTokenizer};
Decoupling global information to compress bitrates, such as TiCodec\,\citep{Ren2024TiCodec} and LSCodec\,\citep{Luo2024LSCodec};
Variable frame lengths, represented by SNAC\,\citep{Hubert2024SNAC}. 
Furthermore, certain studies endeavor to integrate acoustic and semantic tokens to leverage the advantages of both paradigms. 
For instance, SpeechTokenizer\,\citep{zhang2023speechtokenizer} and Mimi\,\citep{defossez2024moshi} adopt a semantic distillation approach, distilling outputs from unsupervised pre-trained models into the first-level codebook. 
In contrast, XCodec\,\citep{Ye2024codecdoesmatter} and XYTokenizer\,\citep{Gong2025XYTokenizer} directly utilize inputs from pre-trained models and constrain the retention of semantic information through semantic reconstruction loss or supervised ASR tasks. 
Additionally, SemantiCodec\,\citep{Liu2024SemantiCodec} and LLM-Codec\,\citep{Yang2024LLMCodec} directly employ fixed strong semantic codebooks.

We summarize our contributions in the following:
\begin{itemize}
  \item We designed a decoupled semantic-acoustic tokenizer architecture. The tokenizer adopts an architecture that combines convolution and Transformer, aiming to better capture low-frequency semantic features and high-frequency acoustic features, which also enables flexible configuration of codebook schemes tailored to specific task requirements.
  \item We verified the effectiveness of multi-stage training, which serves as a key approach to achieving both low bitrates and high audio quality simultaneously. This method also enables the flexible application of individual modules across different scenarios.
  \item We designed a detokenizer architecture to support streaming decoding. Furthermore, by leveraging data optimization strategies and integrating bandwidth extension techniques, the decoder achieves a balance of low complexity, low latency and high audio quality.
\end{itemize}

\section{Thoughts of Design}
\label{thoughts}

\subsection{Key Design Challenges in Text-Speech Multimodal Modeling}

\subsubsection{Synergizing Speech Modal Understanding and Generative Efficiency}
Understanding in text-speech multimodal systems refers to the ability to establish robust cross-modal associations between text and speech, leveraging the rich knowledge embedded in textual corpora while minimizing reliance on large-scale aligned training data. 
While acoustic-only tokenization frameworks have been validated in prior work\,\citep{si2024hertz}, their practicality is hindered by exorbitant data demands. 
This underscores the need for semantically enriched intermediaries—such as Chain-of-Thought (CoT) mechanisms—as bridging layers to facilitate efficient knowledge transfer between text and acoustic modalities.
Complementarily, generative proficiency requires balancing high-fidelity output with strict latency constraints, a cornerstone of real-time conversational systems. 
Recent advances, such as semantic tokenization paired with Flow Matching algorithms, address part of this challenge, but critical limitations persist. Architectures rooted in zero-shot text-to-speech (TTS) paradigms, for instance, exhibit inefficiencies in interactive scenarios due to their narrow focus on speech synthesis.
Moreover, the exclusion of fine-grained acoustic features in such frameworks undermines the generalization benefits of pretraining, limiting their adaptability across diverse tasks.

\subsubsection{Aligning Speech Modal Characteristics with Model Capabilities}
A fundamental disparity between text and speech modalities lies in their semantic information density: text conveys meaning with far fewer tokens, whereas speech tokens—characterized by lower density—impose heavier demands on the language model’s (LM) contextual capacity. 
Specifically, conveying equivalent content via speech requires more tokens, straining the LM’s context window under fixed capacity constraints.
A direct strategy to alleviate this burden is to increase the frame length for speech token modeling, thereby reducing the total number of tokens generated in the time sequence. 
However, this approach improves contextual efficiency at the expense of temporal resolution; without introducing multiple codebooks, it would result in information loss, which in turn impairs reconstruction capabilities.
When considering multi-codebook architecture, acoustic metrics such as reconstruction similarity can be easily improved by increasing the number of codebooks. 
However, we argue that acoustic metrics like reconstruction similarity are not the critical factors. 
In a Speech LLM system, an optimal balance exists between the number of codebooks and the fidelity of reconstruction.
At this balance point, the number of codebooks is friendly to Autoregressive (AR) processes, while a further increase in the number of codebooks would not lead to a perceptible improvement in auditory perception. 
Thus, a core challenge emerges: determining the optimal frame length and number of codebooks in order to balance the contextual processing constraints of hierarchical language models with the requirement to preserve fine-grained information. Achieving this balance is essential for maintaining both model efficiency and output quality.

This clarifies the dual dimensions of the design challenge: (1) coordinating cross-modal understanding and real-time generation, and (2) adapting model design to inherent modality differences, such as information density and frame dynamics.
Together, these sections highlight the interconnected trade-offs that underpin effective text-speech multimodal system design.

\subsection{Model Design}
\subsubsection{Encoder Architecture Design}
In the training pipeline of Speech LLM, speech tokens are typically pre-extracted before training the AR model. 
Consequently, the encoder component of the tokenizer needs not adhere to a streaming design. 
A segment-based encoder, by modeling contextual information across temporal frames, inherently endows each token with bidirectional contextual awareness. 
This architecture naturally facilitates multi-token prediction capabilities, which are particularly advantageous for enhancing the learning efficiency of the backbone AR model. 
Furthermore, frame length constitutes another critical factor. 
Shorter frame lengths are more conducive to reconstruction; however, they are more prone to reaching the upper limit of the context window in AR model.

\subsubsection{Transformer vs. Convolutional Layers}
Convolutional layers, characterized by their inductive bias properties, exhibit superior robustness on unseen data, rendering them well-suited for extracting acoustic features. 
In contrast, transformers excel at capturing long-range dependencies and semantic relationships, making them more appropriate for semantic feature extraction tasks. 
A synergistic integration of these two paradigms can effectively balance local acoustic modeling and global semantic understanding.

\subsubsection{Decoder Architecture Design}
The decoder design should prioritize both high-fidelity speech generation and low-latency inference. 
High-fidelity generation should be distinguished from mere reconstruction similarity: it emphasizes the preservation of semantic and prosodic information while reconstructing clear, natural-sounding speech, rather than merely reconstructing the input audio. 
Low latency is achieved through a streaming architecture that ensures consistency between training and inference phases, requiring minimal future context to generate high-quality speech outputs.

\subsection{Codebook Settings}
\subsubsection{Codebook Numbers}
In the semantic distillation framework, semantic and acoustic information share a multi-codebook structure. 
Our experimental observations reveal that the continuous injection of semantic information into deeper codebooks compromises the model’s ability to capture and reconstruct fine-grained audio details.
Consequently, we restrict semantic encoding to a single codebook, while adopting a residual codebook architecture exclusively for acoustic codebooks to flexibly accommodate diverse requirement scenarios.
This design introduces an additional consideration: determining the optimal number of acoustic codebooks.
Increasing the number of codebooks generally enhances the information capacity and reconstruction fidelity. 
However, for AR models, excessive codebooks can lead to exponential growth in inference complexity.
While techniques such as RQ-Transformer\,\citep{defossez2024moshi} or Delay Pattern\,\citep{Jade2023MusicGen} have been proposed to mitigate this issue, they inevitably introduce additional computational overhead and latency. 
Therefore, determining an optimal codebook configuration tailored to specific application requirements is imperative for achieving an efficient balance between model expressiveness and real-time performance.

\subsubsection{Codebook Size}
An AR model can generate sufficiently diverse logits in a single inference step, thus justifying the need to scale up the codebook size to match the model's capabilities. 
However, an excessively large codebook in vector quantization tends to induce codebook collapse, a phenomenon attributed to multiple factors such as insufficient encoder capacity, suboptimal data distribution, and excessively high search dimensionality.
Common strategies to mitigate codebook collapse include scalar quantization methods like LFQ\,\citep{Yu2023lfq} and FSQ\,\citep{Mentzer2023fsq}. 
Nevertheless, such quantization approaches rely heavily on a powerful decoder to interpret the information encapsulated within the codebook, making them incompatible with streaming decoder-based reconstruction. 
Therefore, there is a pressing need to design a codebook that concurrently accommodates the requirements of both AR modeling and decoder functionality.

\section{Model}
\label{model}
\begin{figure}[h]
  \centering
  \includegraphics[width=1.0\textwidth]{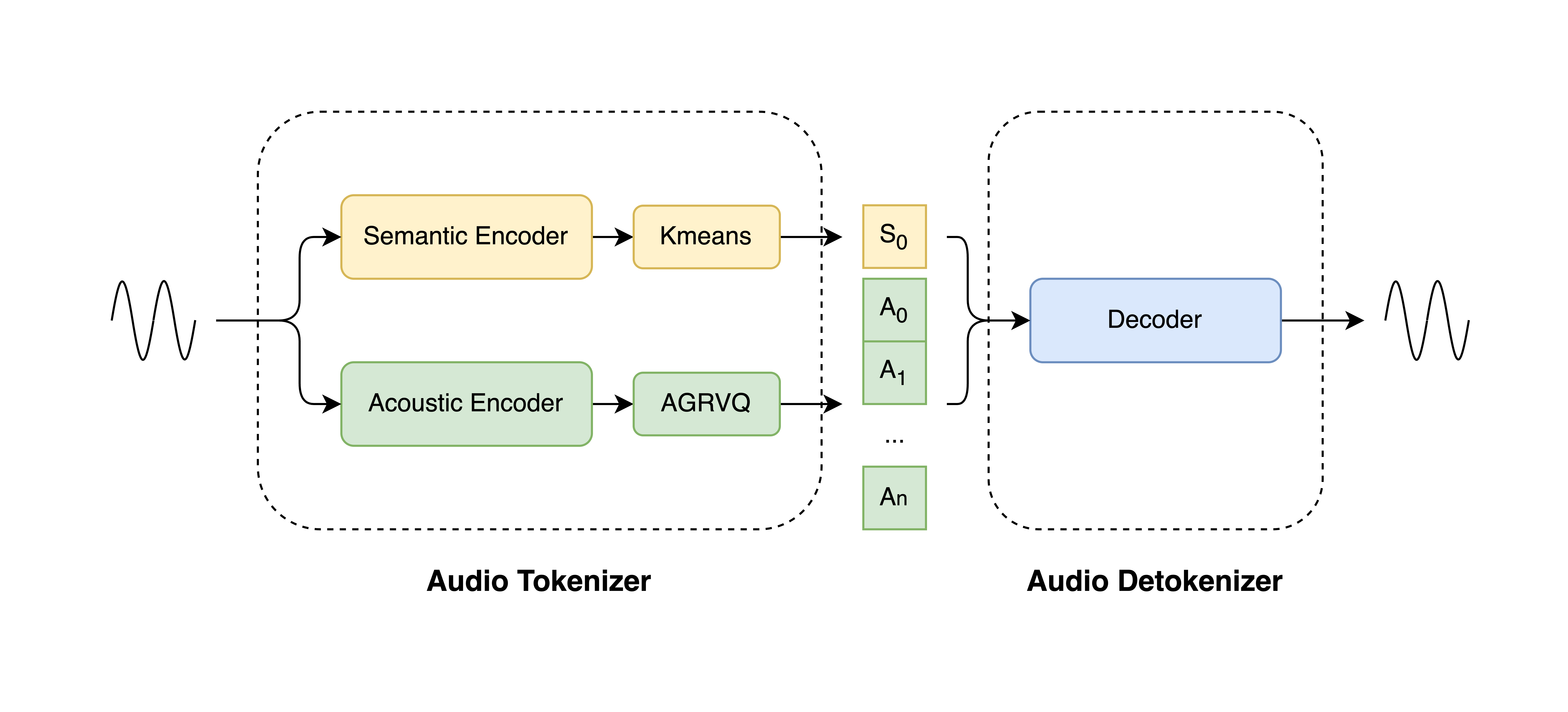}
  \caption{Architecture of LongCat-Audio-Codec}
  \label{fig:codec}
\end{figure}

Guided by the aforementioned design thoughts, we proposed LongCat-Audio-Codec. 
As illustrated by Figure\,\ref{fig:codec}, LongCat-Audio-Codec is built upon two core elements: (1) the Audio Tokenizer (left) for speech signal discretization and (2) the Audio Detokenizer (right) for audio waveform reconstruction which are seamlessly integrated via Quantization modules.

\subsection{Tokenizer}

The tokenization strategy presents a fundamental trade-off: semantic tokens, while effective for capturing linguistic meaning, tend to discard valuable acoustic details, thereby restricting the model's ability to learn intricate non-semantic patterns from extensive speech corpora. 
However, purely acoustic tokens operate at a lower level of abstraction, potentially impeding the model's capacity to effectively process and utilize semantic information. 
Our preliminary experiments demonstrate this limitation, showing that models trained solely on acoustic tokens struggle with basic speech continuation tasks.

Building upon these findings, we introduce a semantic-acoustic tokenization framework. Our architecture implements specialized encoders for independent feature extraction: a semantic encoder for linguistic information capture and an acoustic encoder for detailed speech characteristic preservation. 
This dual-path approach facilitates more robust and comprehensive speech representation learning.

\subsubsection{Semantic Tokenizer}
\label{semantic_tokenizer}

\begin{figure}[h]
  \centering
  \includegraphics[width=0.7\textwidth]{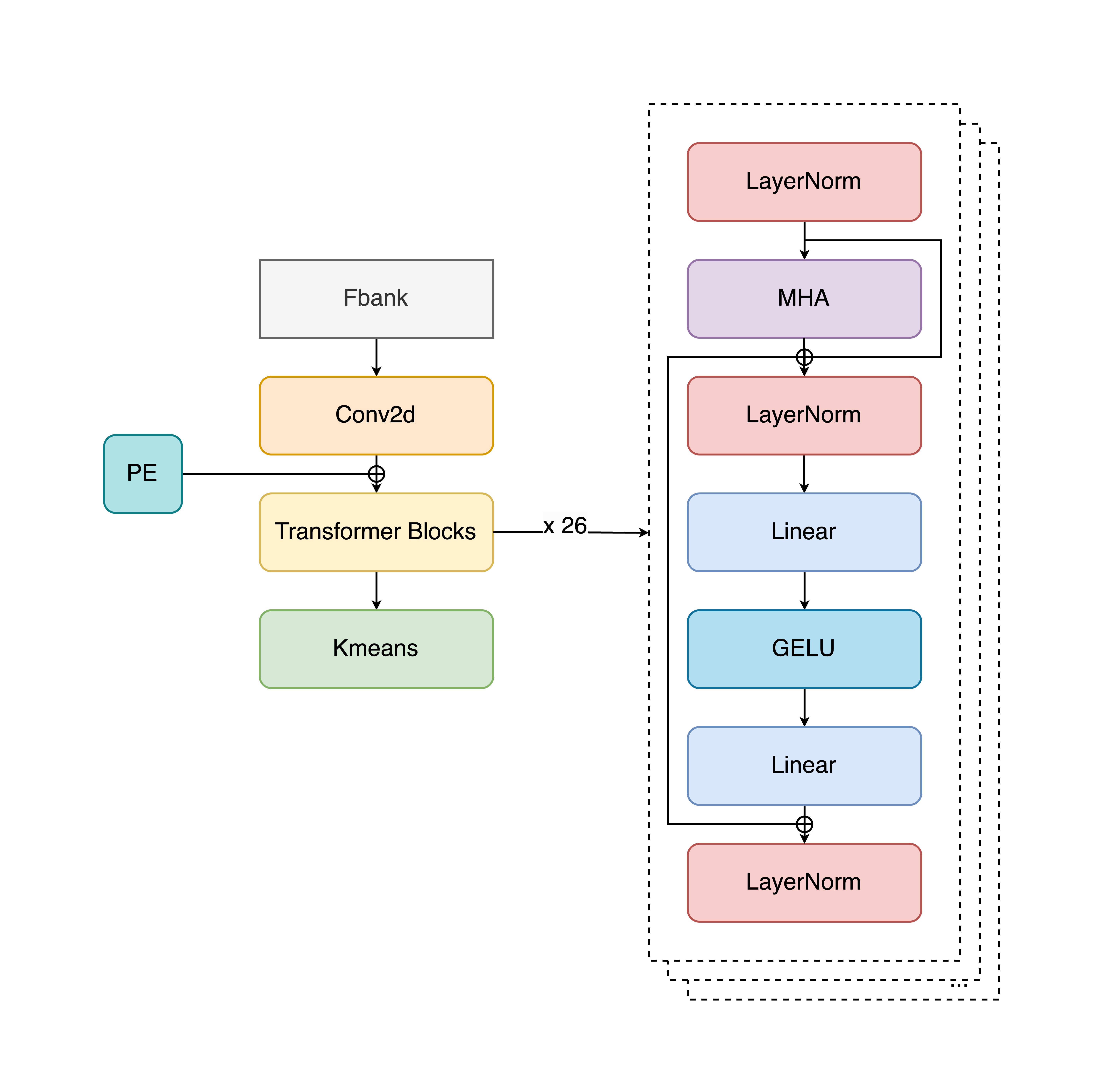}
  \caption{Architecture of semantic encoder}
  \label{fig:semantic_encoder}
\end{figure}

For semantic token extraction, we leverage hidden-layer embeddings from a bi-directional transformer model for Kmeans clustering. 
The transformer processes fbank features with a temporal resolution of $10$ms stride and $25$ms window size. 
A two-layered Conv2d architecture transforms these fbank features into frames of $60$ms, which subsequently serve as input to the transformer blocks. 
Model details can be found in Figure\,\ref{fig:semantic_encoder}

\subsubsection{Acoustic Tokenizer}
\label{acoustic_tokenizer}

\paragraph{Acoustic Encoder}
Considering that the semantic encoder has already performed feature clustering based on Transformer to obtain low-frequency and high-dimensional feature information, the missing high-frequency feature information should be supplemented by non-attention mechanisms. 
We modified the acoustic encoder provided by the method in \citep{kumar2024dac} to have the same frame rate configuration as that of the semantic encoder. 

\paragraph{Acoustic Quantizer}

\begin{figure}[h]
  \centering
  \includegraphics[width=0.6\textwidth]{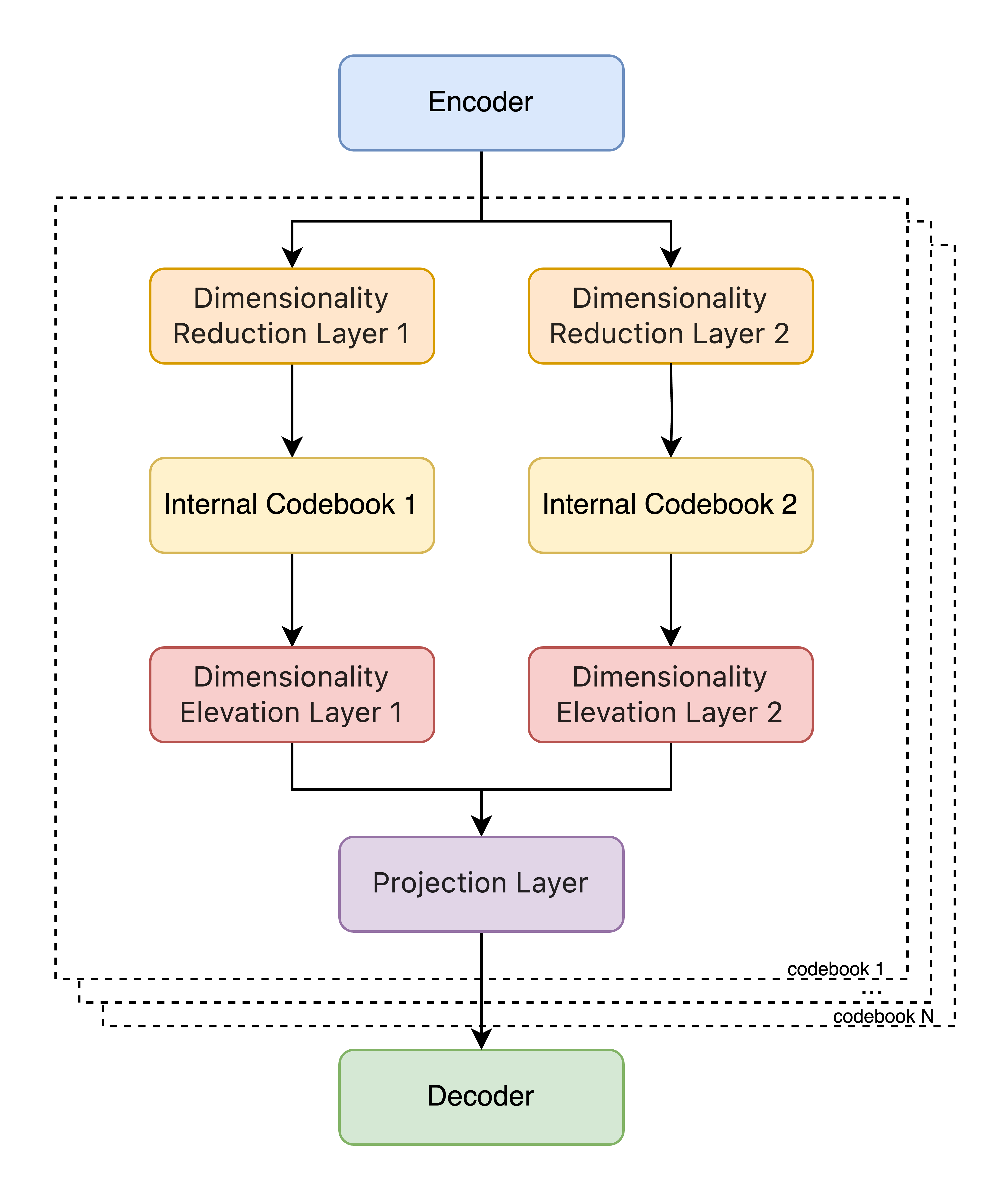}
  \caption{Architecture of AGRVQ}
  \label{fig:agrvq}
\end{figure}

Inspired by \citep{Yang2023hificodec}, we designed an Adaptive Grouped Residual Vector Quantization (AGRVQ) to enhance the training stability of the quantizer under a large codebook size and improve the reconstruction quality of the streaming decoder. 
Compared with direct split grouping, adaptive grouping can effectively improve training stability.
To decouple the constraints of the codebook from those of the latent of the acoustic encoder, we utilized two distinct projection layers to perform basis - change operations on the latent output of the acoustic encoder respectively. 
Meanwhile, to further enhance the stability of the codebook and reduce the search cost, we carried out dimensionality-reduction operations on these latents. 
The fundamental workflow of AGRVQ is illustrated by Figure\,\ref{fig:agrvq}.
Each of the two new latents is encoded using an independent internal codebook, with the codebook size of each codebook being $90$ in current version of LongCat-Audio-Codec. 
These two internal codebooks are combined to form an equivalent acoustic codebook with a size of $90 \times 90 = 8100$. 
We employed 3 acoustic codebooks to construct a 4-codebook codec, as our analysis indicated that 4 codebooks are sufficient to capture speech information effectively. Further details are provided in Appendix A.

\subsection{Detokenizer}

\begin{figure}[h]
  \centering
  \includegraphics[width=0.6\textwidth]{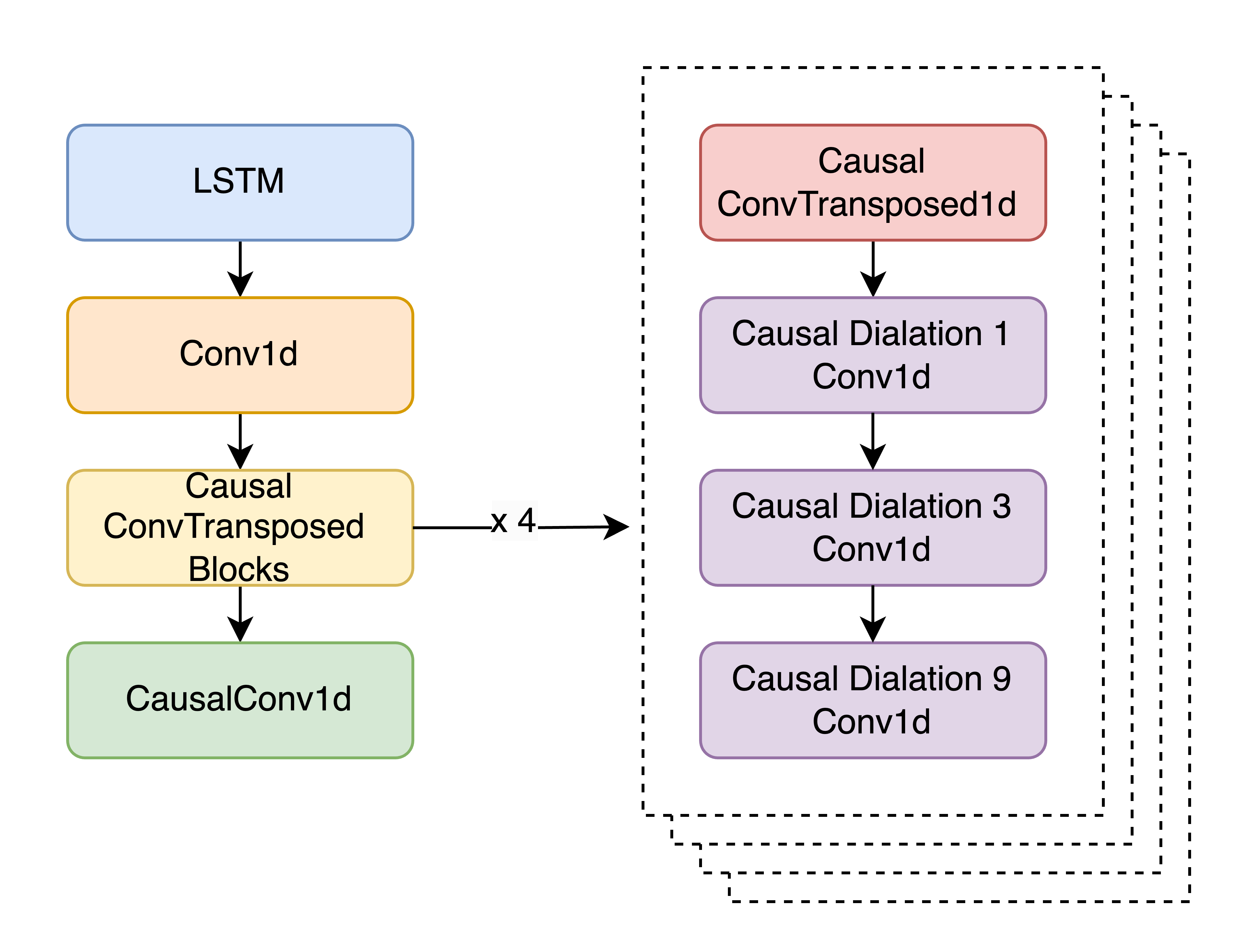}
  \caption{Architecture of detokenizer (decoder)}
  \label{fig:audio_decoder}
\end{figure}

To enable semantic information to participate in reconstruction, the detokenizer(in the following context, "decoder" and "detokenizer" refer to the same concept) needs to take both semantic latents and acoustic latents as inputs. 
Since semantic latents are pre-trained content, the trainable acoustic information will naturally compensate for the missing information in semantic information that is required for reconstruction. We found that the decoder reconstructs audio by referencing both semantic latents and acoustic latents, with a greater emphasis on the acoustic component.
As illustrated by Figure\,\ref{fig:audio_decoder}, the audio detokenizer was designed to supporting stream decoding, composed of LSTM, convolution layers and causal convolution transposed layers, with a look-ahead of 3 frames (corresponding to latency of $180$ms). 
This structure features training-inference consistent streaming inference capability, eliminating the need for additional mask designs required by diffusion-based detokenizers, while demonstrating significant advantages in terms of computational complexity.

\section{Training and Evaluation}
\label{eval}

\subsection{Training of Semantic Encoder}

Our semantic training pipeline begins with unsupervised pre-training using the BEST-RQ\,\citep{chiu2022bestrq} method on in-house speech data, followed by supervised fine-tuning for ASR tasks of Chinese and English with CTC loss\,\citep{graves2006connectionist}. 
The complete semantic encoder contains approximately $500$M parameters.
We subsequently conducted a comprehensive comparison of K-means clustering results using embeddings extracted from various hidden layers of both models.
Downstream task evaluations, as listed in Table\,\ref{tab:performance_metrics}, demonstrate that the last layer of the CTC fine-tuned model contains the most semantically rich representations. 
The proposed semantic-acoustic splited architecture enables a specialized division of labor: while the acoustic codebook effectively captures detailed acoustic information, we strategically optimized our semantic tokens to focus exclusively on preserving rich semantic content, without concern for potential information loss in other domains.
Consequently, we selected the last hidden layer representations from the CTC fine-tuned model to enhance semantic token quality.

The optimized semantic tokenizer configuration features an $8192$-entry codebook and operates at a frame rate of approximately $16.67$Hz (equating to $60$ms temporal resolution per frame), achieving an optimal balance between semantic richness and computational efficiency.

\subsection{Training of Acoustic Encoder and Decoder}

\begin{figure}[h]
  \centering
  \includegraphics[width=0.8\textwidth]{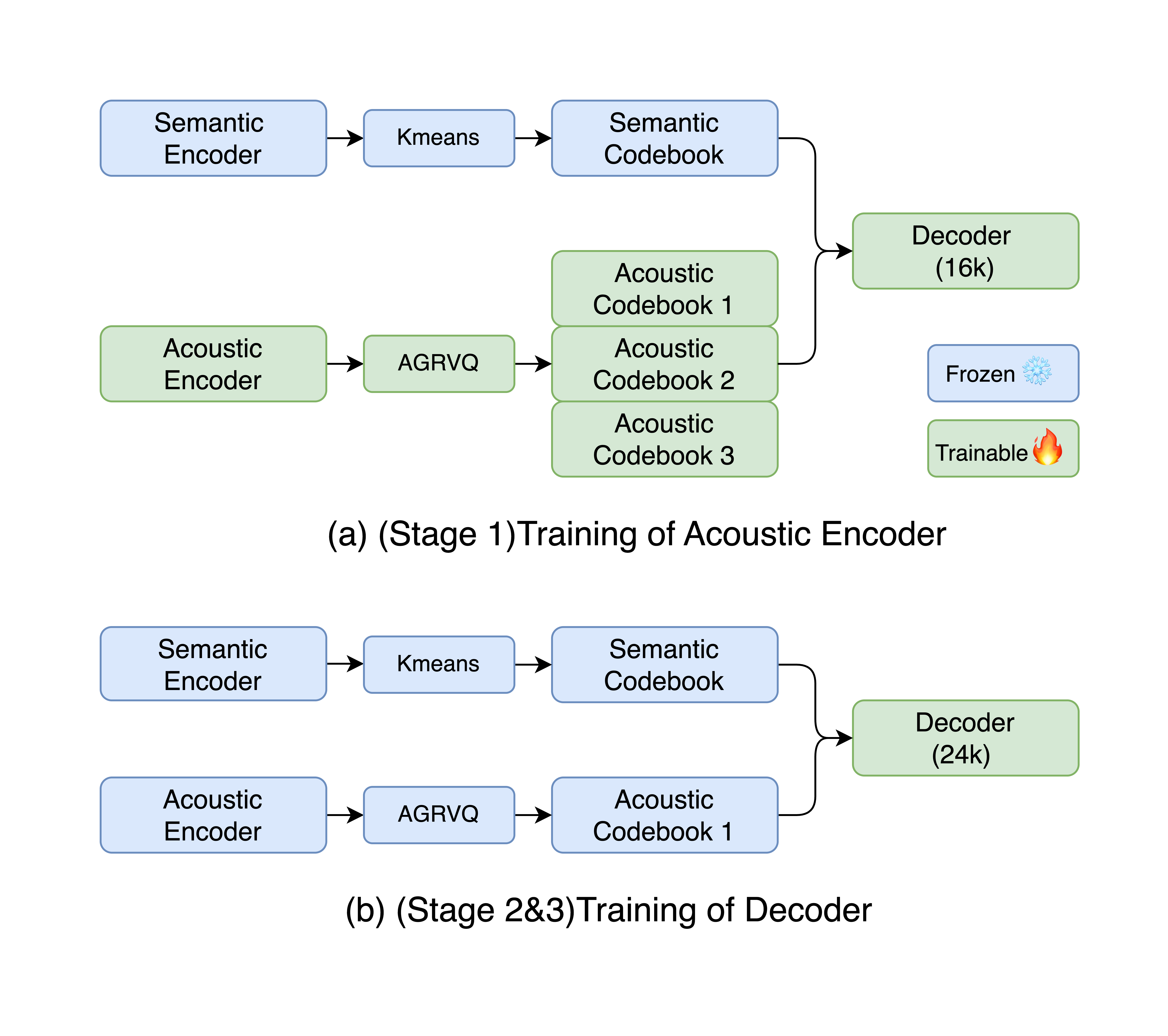}
  \caption{Acoustic training pipeline}
  \label{fig:tokenizer_pipeline}
\end{figure}

The training objective of the acoustic encoder is to complement the missing reconstruction information in the semantic component, rather than being independently dedicated to reconstruction. 
After obtaining the semantic tokenizer, we designed a multi-stage training strategy for training the acoustic component as illustrated in the Figure\,\ref{fig:tokenizer_pipeline}. 
The objective of Stage 1 is to expose the model to as diverse a range of speech data as possible, thereby ensuring that the tokens modeled by the encoder exhibit a sufficiently uniform distribution and possess adequate robustness during token extraction. 
For Stage 2, leveraging the Train-More-Use-Less(TMUL) technique, the goal is to select an appropriate number of codebooks based on downstream scenario requirements, and train the decoder on specific data to enhance the reconstruction stability, audio quality, and other attributes associated with the designated codebooks. 
Stage 3, an optional step, further specializes the decoder for specific speakers and is primarily used in downstream scenarios with a limited number of speakers.

\subsubsection{Stage 1: Encoder Pretrain}
To enhance the robustness of the Encoder, it is essential for the Encoder to be exposed to a sufficiently rich set of acoustic environments during training. 
We first collect about $500,000$ hours audio data encompassing diverse speech patterns and acoustic environments with sampling rate of $16$kHz (downsampled if necessary). 
We perform uniform sampling on the data to ensure that the observed data distribution is adequately diverse. 
During the training process, we first utilize only the mel loss, and activate the discriminator once the training stabilizes. Notably, the current model design and Stage 1 training strategy are not optimized for reconstruction metrics. While certain designs and strategies capable of improving reconstruction metrics would compromise the preservation of semantic information, Stage 1 prioritizes maximizing intelligibility over acoustic details—considering that audio quality and timbre similarity can be enhanced in Stages 2 and 3.

\subsubsection{Stage 2: Decoder Pretrain}
During the decoder training phase, we freeze the encoder and quantizer, select the required number of acoustic codebooks (the example in Figure\,\ref{fig:tokenizer_pipeline} uses 2 codebooks: one semantic codebook and one acoustic codebook), and only train the decoder. 
At this stage, the parameter configuration of the decoder is adjusted such that each token decodes $0.06 \times 24000 = 1440$ samples, thereby achieving a sample rate conversion from $16$k to $24$k. We scale up the final decoder parameters to $150$M  (while larger parameter scales were explored, only reconstruction quality gains were observed, but no significant benefits in generation quality). 
The training data used in this phase are all high-quality, consisting of two parts: approximately 1000 hours of high-quality recorded data collected internally, and $250,000$ hours of data processed by an in-house generative enhancement model. 
Training the decoder exclusively on high-quality data endows it with self-repairing capabilities, where issues such as missing frequency bands and damaged spectra can be restored during this phase.

\subsubsection{Stage 3: Decoder SFT (optional)}
Speech LLM typically supports a limited set of speakers. 
To better align with the requirements of Speech LLM, the pre-trained decoder can undergo further few-round SFT on limited number of speakers, aiming to enhance its capability in reconstructing the voice of a particular speaker. 
We have demonstrated emotion-rich demonstrations of 2-codebooks Stage 3 decoder on the open-source page.

\subsection{Evaluation}

\subsubsection{Evaluation of Semantic Tokenizer}
To better evaluate the information carried by semantic tokens, we constructed an evaluation framework consisting of Chinese-English automatic speech recognition(ASR), speaker identification (SID), sound event detection (SED), and emotion recognition (ER). 
Evaluations were conducted under this framework. 
In terms of dataset selection, the AISHELL-1\citep{huang2017aishell} dataset is employed for Chinese ASR tasks, while LibriSpeech\citep{panayotov2015librispeech} is utilized for English ASR tasks.
For speaker recognition, VoxCeleb1\citep{nagrani2017voxceleb} is adopted. AudioSet\citep{gemmeke2017audio} is used for SED and IEMOCAP\citep{busso2008iemocap} is employed for emotion detection. 
We use frozen latents or tokens as features for each task, conduct training with data augmentation on the training sets, and evaluate various metrics on the test sets.

\begin{table}[ht]
    \centering
    \begin{threeparttable}
    \caption{Downstream task performance of BEST-RQ model}
    \label{tab:performance_metrics_bestrq}
    \begin{tabular}{lcccccc}
        \toprule
        \textbf{Model} & \textbf{ASR-ZH} & \textbf{ASR-EN} & \textbf{SID} & \textbf{SED} & \textbf{ER} \\
                       & \textbf{CER$\downarrow$} & \textbf{UER$\downarrow$} & \textbf{ACC$\uparrow$} & \textbf{ACC$\uparrow$} & \textbf{ACC$\uparrow$} \\
        \midrule
        \textbf{BEST-RQ Layer 2} & 10.70 & 23.24 & 0.462 & 0.587 & 0.577 \\
        \textbf{BEST-RQ Layer 6} & 7.70 & 15.28 & \textbf{0.807} & \textbf{0.596} & 0.635 \\
        \textbf{BEST-RQ Layer 10} & 6.26 & 11.42 & 0.732 & 0.583 & \textbf{0.666} \\
        \textbf{BEST-RQ Layer 14} & 5.26 & 9.32 & 0.514 & 0.564 & 0.644 \\
        \textbf{BEST-RQ Layer 18} & 4.42 & 8.34 & 0.347 & 0.555 & 0.635 \\
        \textbf{BEST-RQ Layer 22} & \textbf{4.03} & \textbf{8.13} & 0.259 & 0.545 & 0.634 \\
        \textbf{BEST-RQ Layer 26-last} & 4.42 & 9.13 & 0.128 & 0.549 & 0.629 \\
        \bottomrule
    \end{tabular}
    
    \end{threeparttable}
\end{table}

\begin{table}[ht]
    \centering
    \begin{threeparttable}
    \caption{Downstream task performance of ASR model}
    \label{tab:performance_metrics_asr}
    \begin{tabular}{lcccccc}
        \toprule
        \textbf{Model} & \textbf{ASR-ZH} & \textbf{ASR-EN} & \textbf{SID} & \textbf{SED} & \textbf{ER} \\
                       & \textbf{CER$\downarrow$} & \textbf{UER$\downarrow$} & \textbf{ACC$\uparrow$} & \textbf{ACC$\uparrow$} & \textbf{ACC$\uparrow$} \\
        \midrule
        \textbf{ASR Layer 2} & 10.71 & 24.96 & 0.480 & 0.589 & 0.594 \\
        \textbf{ASR Layer 6} & 8.56 & 18.03 & 0.716 & \textbf{0.614} & 0.620 \\
        \textbf{ASR Layer 10} & 7.11 & 14.94 & 0.876 & \textbf{0.614} & 0.654 \\
        \textbf{ASR Layer 14} & 5.99 & 11.77 & \textbf{0.891} & 0.611 & \textbf{0.680} \\
        \textbf{ASR Layer 18} & 5.24 & 9.09 & 0.788 & 0.600 & 0.650 \\
        \textbf{ASR Layer 22} & 4.00 & 6.29 & 0.526 & 0.580 & 0.677 \\
        \textbf{ASR Layer 26-last} & \textbf{2.91} & \textbf{4.09} & 0.274 & 0.572 & 0.669 \\
        \bottomrule
    \end{tabular}
    
    \end{threeparttable}
\end{table}

\begin{table}[ht]
    \centering
    \begin{threeparttable}
    \caption{Comparative results of continuous/discrete}
    \label{tab:performance_metrics}
    \begin{tabular}{lcccccc}
        \toprule
        \textbf{Model} & \textbf{ASR-ZH} & \textbf{ASR-EN} & \textbf{SID} & \textbf{SED} & \textbf{ER} \\
                       & \textbf{CER$\downarrow$} & \textbf{UER$\downarrow$} & \textbf{ACC$\uparrow$} & \textbf{ACC$\uparrow$} & \textbf{ACC$\uparrow$} \\
        \midrule
        \textbf{BEST-RQ Layer 22} & 4.03 & 8.13 & 0.259 & 0.545 & 0.634 \\
        %\midrule
        \textbf{BEST-RQ Layer 22 (Q)}\textsuperscript{*} & 6.89 & 11.76 & 0.090 & 0.430 & 0.616 \\
        %\midrule
        \textbf{ASR Layer 26} & 2.91 & 4.09 & 0.274 & 0.572 & 0.669 \\
        %\midrule
        \textbf{ASR Layer 26 (Q)}\textsuperscript{*} & 4.91  & 7.31 & 0.051 & 0.418 & 0.575 \\
        \bottomrule
        \end{tabular}
        
        \begin{tablenotes}
        \small
        \item[*] use the representation after Quantization for evaluation.
        \end{tablenotes}

    \end{threeparttable}
\end{table}

Given the acoustic encoder introduces more acoustic details, so when selecting the semantic part, we pay more attention to the test results of speech recognition. 
According to Table\,\ref{tab:performance_metrics_bestrq} and Table\,\ref{tab:performance_metrics_asr}, we found that the 22nd layer latent of BEST-RQ (we also verified the performance of the 21st and 23rd layers to confirm that the 22nd layer is the best) can obtain optimal results from the speech recognition test, while the last layer latent of the ASR model can achieve better speech recognition results. 
At the same time, by comparing the test results of the 22nd layer latent of BEST-RQ and the last layer latent of the ASR model in SID, SED, and ER, the ASR model can achieve relatively better results; this means that starting from the latents of the ASR model, we can obtain better semantic representations. 
We further applied k-means to the 22nd layer of BEST-RQ and the last layer of the ASR model to uniformly sample the latents and form tokens.
We then evaluated these discretized tokens within the evaluation framework. 
The evaluation results Table\,\ref{tab:performance_metrics} show that the discrete tokens based on the ASR model still hold an advantage in the speech recognition task.

\subsubsection{Train-More-Use-Less (TMUL)}
As results shown in Table\,\ref{tab:tmul}, we have observed that using N codebooks selected from a pre-trained M-codebook codec (where N < M) yields a lower average acoustic error compared to directly training an N-codebook codec.

\begin{table}[h!]
\centering
\begin{threeparttable}
\caption{Comparative results of different TMUL configuration on in-house dataset}
\label{tab:tmul}
\begin{tabular}{ccccccc}
\toprule
\textbf{Strategy} & \multicolumn{3}{c}{\textbf{ZH}} & \multicolumn{3}{c}{\textbf{EN}} \\
\cmidrule(lr){2-4} \cmidrule(lr){5-7}
& \textbf{STOI↑}
 & \textbf{MR-STFT$\downarrow$} & \textbf{MR-MEL$\downarrow$} & \textbf{STOI$\uparrow$} & \textbf{MR-STFT$\downarrow$} & \textbf{MR-MEL$\downarrow$} \\
\midrule
train-1-use-1 & 0.734 & 0.138 & 0.456 & 0.699 & 0.142 & 0.378 \\
train-3-use-1 & \textbf{0.736} & \textbf{0.135} & \textbf{0.444} & \textbf{0.712} & \textbf{0.137} & \textbf{0.365} \\
\bottomrule
\end{tabular}

\begin{tablenotes}
\small
\item MR-STFT(MEL): multi-resolution STFT(MEL) distance.
\end{tablenotes}

\end{threeparttable}
\end{table}

This phenomenon can be explained as follows: When training a single codebook, all information required for reconstruction must be compressed into that sole codebook, which causes some outliers to encroach upon the representational space of mainstream information. 
In contrast, during multi-codebook training, the first-layer codebooks can capture the most significant information on average, thereby resulting in a lower average reconstruction error.
Notably, we have also noticed that this strategy may increase the recognition error rate for a small number of audio cases, which further validates our conjecture.

\subsubsection{Evaluation of Stage 1}
Follow \citep{Guo2025DiscreteTokens}, We use LibriTTS\citep{zen2019libritts} testset-B for evaluations. 
We evaluate reconstruction performance of stage 1 ($16$khz pretrained encoder and decoder) using the following metrics:
\begin{itemize}
\item \textbf{Word Error Rate (WER) ($\downarrow$)}: Quantifies the content intelligibility of reconstructed speech by measuring discrepancies between ground-truth transcripts and outputs decoded by an automatic speech recognition (ASR) system. For this study, we employ the NeMo-ASR7 model\footnote{\url{https://github.com/NVIDIA-NeMo/NeMo}} for transcription.

\item \textbf{Gross Pitch Error (GPE) ($\downarrow$)}: Assesses the relative percentage error in pitch contour alignment between reconstructed speech and the ground-truth audio.

\item \textbf{Perceptual Evaluation of Speech Quality (PESQ) ($\uparrow$)}: A standardized metric that objectively assesses speech quality by comparing degraded speech with a reference

\item \textbf{Short-Time Objective Intelligibility (STOI) ($\uparrow$)}: Measures speech intelligibility, quantifying how well the original message is preserved in degraded speech.

\item \textbf{Speaker Embedding Cosine Similarity (SECS) ($\uparrow$)}: Computes the cosine similarity between speaker embeddings derived from reconstructed speech and ground-truth audio, using a pre-trained speaker verification model\footnote{\url{https://github.com/resemble-ai/Resemblyzer}} for embedding extraction.
\end{itemize}

We conducted a grouped comparison between codecs carrying semantic information and pure acoustic codecs. 
Since the introduction of semantic information may affect acoustic reconstruction metrics, establishing two distinct groups is considered a fair approach. 
The semantic codecs here incorporate semantics through means including semantic distillation, CTC loss supervision, semantic reconstruction loss constraints, and direct utilization of semantic tokens for reconstruction.
We grouped the tests based on the bitrate configurations of LongCat-Audio-Codec ($0.43$kbps, $0.65$kbps, $0.87$kbps).
Efforts were made to ensure that LongCat-Audio-Codec operates at the lowest bitrate within each group as far as possible, so as to avoid unfair conditions to other codecs.
Overall, four comparison groups were established: $>2$kbps, $0.85\sim2$kbps, $0.65\sim0.85$kbps, and $<0.65$kbps.
Comparative results demonstrate that by adhering to the Stage 1 training objective of prioritizing high intelligibility, LongCat-Audio-Codec achieves a distinct intelligibility advantage across all comparison groups. Regarding acoustic reconstruction metrics, LongCat-Audio-Codec also outperforms other semantic codecs and exhibits sufficient competitiveness when compared with acoustic codecs.

\paragraph*{Comparation with Semantic Codec}

\begin{table*}[ht]
\centering
\caption{Comparative results of codec with semantic information}
\label{tokenizer_results_semantic}
\adjustbox{max width=\textwidth}{
\begin{tabular}{lccccccc}
\toprule
\multirow{2}{*}{\textbf{\makecell{Model\\(version)}}} & 
\multirow{2}{*}{\textbf{\makecell{Frame Rate\\(Hz)}}} & 
\multirow{2}{*}{\textbf{\makecell{Bitrate\\(kbps)}}} & 
\multicolumn{5}{c}{\textbf{Reconstruction}} \\
& & & \textbf{WER$\downarrow$} & \textbf{GPE$\downarrow$} & \textbf{PESQ$\uparrow$} & \textbf{STOI$\uparrow$} & \textbf{SECS$\uparrow$} \\
\midrule
\multicolumn{8}{c}{\textbf{reference}} \\
\midrule
Ground Truth Recording & \centering - & - & 1.16 & 0.00 & 4.50 & 1.000 & 1.000 \\
Mel + BigVGAN (100 band) & \centering - & - & 1.18 & 0.88 & 4.30 & 0.995 & 0.997 \\
\midrule
\multicolumn{8}{c}{\textbf{>2 kbps}} \\
\midrule
FACodec (with detail codes)\,\citep{Ju2024FACodec} & \centering 80 & 4.80 & \underline{1.37} & \textbf{1.02} & \textbf{2.91} & \textbf{0.954} & 0.971 \\
X-Codec (HuBERT LibriSpeech)\,\citep{Ye2024codecdoesmatter} & \centering 50 & 4.00 & \textbf{1.27} & 1.49 & \underline{2.82} & 0.905 & 0.971 \\
SpeechTokenizer\,\citep{zhang2023speechtokenizer} & \centering 50 & 4.00 & 1.47 & \underline{1.20} & 2.60 & \underline{0.930} & 0.972 \\
\midrule
\multicolumn{8}{c}{\textbf{0.85-2 kbps}} \\
\midrule
Mimi\,\citep{defossez2024moshi} & \centering 12.5 & 1.10 & \underline{2.44} & \underline{1.68} & \underline{2.27} & \underline{0.917} & \underline{0.938} \\
LLM-Codec\,\citep{Yang2024LLMCodec} & \centering 8.33+16.67+33.33 & 0.85 & 6.25 & 1.86 & 1.82 & 0.879 & 0.919 \\
LongCat-Audio-Codec (4 codebooks) & \centering 16.7 & 0.87 & \textbf{1.48} & \textbf{1.65} & \textbf{2.30} & \textbf{0.921} & \textbf{0.942} \\ 
\midrule
\multicolumn{8}{c}{\textbf{0.65-0.85 kbps}} \\
\midrule
SemantiCodec ($F=25$Hz)\,\citep{Liu2024SemantiCodec} & \centering 25 & 0.70 & \underline{3.44} & \underline{2.28} & \underline{1.75} & \underline{0.866} & \textbf{0.930} \\
LongCat-Audio-Codec (3 codebooks) & \centering 16.7 & 0.65 & \textbf{1.70} & \textbf{1.86} & \textbf{2.01} & \textbf{0.900} & \underline{0.925} \\ 
\midrule

\multicolumn{8}{c}{\textbf{<0.65 kbps}} \\
\midrule
$\mathcal{S}^3$ Tokenizer ($F=50$Hz)\,\citep{du2024cosyvoice}\textsuperscript{*} & \centering 50 & 0.60 & \underline{2.12} & - & - & 0.673 & - \\
LSCodec ($F=50$Hz)\,\citep{Luo2024LSCodec}\textsuperscript{*} & \centering 50 & 0.45 & 3.33 & - & - & \underline{0.688} & - \\
LongCat-Audio-Codec (2 codebooks) & \centering 16.7 & 0.43 & \textbf{2.10} & 3.69 & 1.47 & \textbf{0.839} & 0.862 \\

\bottomrule
\end{tabular} 

}
\begin{tablenotes}
    \small
    \item  \textbf{bold} indicates the best performance in the group; \underline{underlined} indicates the second-best.
    \item  \textsuperscript{*} has extra speaker embedding on decoder, only WER and STOI are evaluated.
\end{tablenotes}
\end{table*}

As shown in Table\,\ref{tokenizer_results_semantic}, the proposed LongCat-Audio-Codec exhibits certain favorable characteristics across multiple evaluation dimensions compared to other semantic codecs, with particular potential in balancing low-bitrate efficiency and reconstruction performance.

In the $0.85\sim2$kbps range, LongCat-Audio-Codec (4 codebooks) operates at $0.87$kbps in this range. 
Compared to counterparts (Mimi at $1.10$kbps and LLM-Codec at $0.85$kbps), it demonstrates comprehensive superiority across all evaluated metrics:
It achieves the lowest WER($1.48$) and GPE($1.65$), indicating superior speech intelligibility and speaker-related feature preservation.
It also attains the highest PESQ($2.30$), STOI($0.921$), and SECS ($0.942$), reflecting better overall perceptual quality and consistency.

In the $0.65\sim0.85$kbps range, LongCat-Audio-Codec (3 codebooks) operates at $0.65$kbps here. 
Against SemantiCodec ($0.70$kbps) and $\mathcal{S}^3$ Tokenizer ($0.60$kbps), it shows notable advantages:
It delivers the lowest WER ($1.70$) and GPE ($1.86$), outperforming SemantiCodec (WER: $3.44$; GPE: $2.28$).
It achieves the highest PESQ ($2.01$) and STOI ($0.900$), significantly surpassing SemantiCodec (PESQ: $1.75$; STOI: $0.866$). 
Its SECS (0.925) is among the highest, highlighting strong performance in semantic or subjective consistency.

In the $<0.65$kbps range, LongCat-Audio-Codec (2 codebooks) operates at $0.43$kbps. Compared to $\mathcal{S}^3$ Tokenizer($0.60$kbps) and LSCodec ($0.45$kbps),
it achieves a substantially lower WER (2.10 vs. 2.12 vs. 3.33) and a higher STOI (0.839 vs. 0.673 vs. 0.688), indicating better intelligibility retention under extreme bitrate constraints.

In summary, When compared to semantic codec of similar bitrates, LongCat-Audio-Codec consistently demonstrates strong performance, excelling in both intelligibility (WER, STOI) and reconstruction quality (GPE, PESQ, SECS). 
\paragraph*{Comparation with Acoustic Codec}

\begin{table*}[ht]
\centering
\caption{Comparative results of codec without semantic information}
\label{tokenizer_results_acoustic}
\adjustbox{max width=\textwidth}{
\begin{tabular}{lccccccc}
\toprule
\multirow{2}{*}{\textbf{\makecell{Model\\(version)}}} & 
\multirow{2}{*}{\textbf{\makecell{Frame Rate\\(Hz)}}} & 
\multirow{2}{*}{\textbf{\makecell{Bitrate\\(kbps)}}} & 
\multicolumn{5}{c}{\textbf{Reconstruction}} \\
& & & \textbf{WER$\downarrow$} & \textbf{GPE$\downarrow$} & \textbf{PESQ$\uparrow$} & \textbf{STOI$\uparrow$} & \textbf{SECS$\uparrow$} \\
\midrule
\multicolumn{8}{c}{\textbf{reference}} \\
\midrule
Ground Truth Recording & \centering - & - & 1.16 & 0.00 & 4.50 & 1.000 & 1.000 \\
Mel + BigVGAN (100 band) & \centering - & - & 1.18 & 0.88 & 4.30 & 0.995 & 0.997 \\
\midrule
\multicolumn{8}{c}{\textbf{>2 kbps}} \\
\midrule
EnCodec ($Q=8$)\,\citep{defossez2022encodec} & \centering 75 & 6.00 & \underline{1.53} & \underline{1.33} & \underline{2.83} & \underline{0.946} & \underline{0.979} \\
DAC (24kHz, $Q=8$)\,\citep{kumar2024dac} & \centering 75 & 6.00 &\textbf{1.34} & \textbf{0.93} & \textbf{3.52} & \textbf{0.958} &\textbf{ 0.982} \\
\midrule
\multicolumn{8}{c}{\textbf{0.85-2 kbps}} \\
\midrule
TiCodec ($Q=2$)\,\citep{Ren2024TiCodec} & \centering 75 & 1.50 & 3.31 & \underline{1.51} & 2.00 & 0.898 & 0.905 \\
SNAC (24kHz)\,\citep{Hubert2024SNAC} & \centering 12+23+47 & 0.98 & \underline{2.25} & \textbf{1.48} & 2.23 & 0.914 & \underline{0.952} \\
WavTokenizer (Small)\,\citep{Ji2024WavTokenizer} & \centering 75 & 0.90 & 2.45 & 1.63 & \textbf{2.47} & \textbf{0.925} & \textbf{0.960} \\
LongCat-Audio-Codec (4 codebooks) & \centering 16.7 & 0.87 & \textbf{1.48} & 1.65 & \underline{2.30} & \underline{0.921} & 0.942 \\ 
\midrule
\multicolumn{8}{c}{\textbf{0.65-0.85 kbps}} \\
\midrule
Stable-Codec ($Q=12$)\,\citep{Julian2024StableCodec} & \centering 25 & 0.70 & \underline{4.94} & \textbf{1.73} & \textbf{2.16} & \textbf{0.917} & \underline{0.889} \\
LongCat-Audio-Codec (3 codebooks) & \centering 16.7 & 0.65 & \textbf{1.70} & \underline{1.86} & \underline{2.01} & \underline{0.900} & \textbf{0.925} \\ 
\midrule
\multicolumn{8}{c}{\textbf{<0.65 kbps}} \\
\midrule
LongCat-Audio-Codec (2 codebooks) & \centering 16.7 & 0.43 & 2.10 & 3.69 & 1.47 & 0.839 & 0.862 \\

\bottomrule
\end{tabular} 

}
\begin{tablenotes}
  \small
  \item  \textbf{bold} indicates the best performance in the group; \underline{underlined} indicates the second-best.
\end{tablenotes}
\end{table*}
As shown in Table\,\ref{tokenizer_results_acoustic}, compared with acoustic codecs of similar bitrates, LongCat-Audio-Codec not only maintains its advantage in recognition accuracy but also achieves comparable results in multiple acoustic reconstruction metrics. 
This demonstrates that it can still retain competitive acoustic reconstruction capabilities even with the introduction of semantic information and less framerate.
In the $0.85\sim2$ kbps range, LongCat-Audio-Codec (4 codebooks) achieves a WER of 1.48, which is significantly lower than all acoustic codec in this range:
It outperforms TiCodec (1.50 kbps, WER=3.31), SNAC ($0.98$kbps, WER=2.25), and WavTokenizer ($0.90$kbps, WER=2.45) by a substantial margin.
While its PESQ (2.23) and STOI (0.921) are slightly trailing WavTokenizer, but maintains competitive performance relative to SNAC (PESQ=2.23, STOI=0.914) and TiCodec (PESQ=2.00, STOI=0.898).
In the $0.65\sim0.85$ kbps range, LongCat-Audio-Codec (3 codebooks) again excels in WER(1.70), outperforming Stable-Codec (4.94) by a significant margin.

\paragraph*{Trend of Metric Changes in LongCat-Audio-Codec Across Different Bitrates}
LongCat-Audio-Codec achieves gradual bitrate adjustment from $0.43$kbps to $0.87$kbps by modifying the number of codebooks (2/3/4). 
Its core metrics exhibit distinct and consistent trends with increasing bitrates:  

\begin{itemize}
    \item \textbf{WER ($\downarrow$)}: Consistently decreases with increasing bitrate (from 2.10 to 1.48, a 29.5\% reduction), indicating improved speech recognition accuracy.  
    \item \textbf{GPE ($\downarrow$)}: Significantly decreases with increasing bitrate (from 3.69 to 1.65, a 55.3\% reduction), reflecting substantial reduction in prosodic errors during speech generation.  
    \item \textbf{PESQ ($\uparrow$)}: Steadily increases with increasing bitrate (from 1.47 to 2.30, a 56.5\% improvement), demonstrating gradual optimization of subjective speech quality (clarity and naturalness).  
    \item \textbf{STOI ($\uparrow$)}: Shows a steady increase with increasing bitrate (from 0.839 to 0.921, a 9.8\% improvement), with speech intelligibility approaching the level of high-bitrate models.  
    \item \textbf{SECS ($\uparrow$)}: Exhibits a slight increase with increasing bitrate (from 0.862 to 0.942, a 9.3\% improvement), with structural similarity of speech signals gradually approaching the original.  
\end{itemize}

\paragraph*{Flexibility in Acoustic Token Usage for Diverse Reconstruction - Generation Needs} 
A key advantage of LongCat-Audio-Codec lies in it can choose acoustic tokens flexibly while ensuring semantic information remains intact, enabling adaptation to diverse reconstruction and generation demands. 
The 2 codebooks configuration, with its ultra-low 0.43 kbps bitrate, is well-suited for scenarios prioritizing efficiency—such as real time generation, where a balance between speed and basic quality is critical(An excessive number of codebooks necessitates the introduction of strategies such as RQ-Transformer or delay-pattern, which leads to an increase in latency). 
For applications requiring higher fidelity (e.g., precise speech reconstruction), increasing to 2 or 3 acoustic codebooks progressively enhances reconstruction performance, as shown in the stepwise improvements. 
This scalability allows LongCat-Audio-Codec to cater to both lightweight generation tasks and high quality reconstruction needs through targeted adjustment of acoustic codebook count.

\subsubsection{Evaluation of Stage 2 and Stage 3}
As described earlier, Stages 2 and 3 primarily involve retraining the decoder. 
During this process, incorporating specific speakers into the training set enables an enhancement in timbre similarity. 
Additionally, by controlling the audio quality used in training, the decoder is inclined to "translate" tokens into high-quality audio, thereby improving the overall auditory experience of the generated audio. 
Notably, high timbre similarity and high-quality speech can be generated even with the minimum number of codebooks.
All tests in this section are conducted under the minimum codebook configuration (1 semantic codebook + 1 acoustic codebook) to validate the effectiveness of this approach.
\paragraph*{Speaker Similarity}
We conducted an overall speaker similarity evaluation using an in-house dataset featuring a single speaker with rich emotional expressions. 
This evaluation encompasses not only timbre but also the emotions and prosody embedded in the speech. 
Notably, this dataset was not included in the training set for Stage 1, meaning the Encoder had no prior exposure to either the dataset or the specific speaker. 
During Stage 2, this speaker was mixed with a large number of other speakers for high-quality audio training. 
In Stage 3, only few speakers(around 20 speakers, inlcuding the choosen speaker) was used for SFT.
To comprehensively evaluate similarity that incorporates emotions and prosody—rather than focusing solely on timbre—we employed CAM++\,\citep{Wang2023cam++}\footnote{\url{https://modelscope.cn/models/iic/speech_campplus_sv_zh-cn_16k-common/summary}} for assessment, as our findings indicate that the embeddings generated by CAM++ also encapsulate the required emotional and prosodic information.

\begin{figure}[h]
  \centering
  \includegraphics[width=0.8\textwidth]{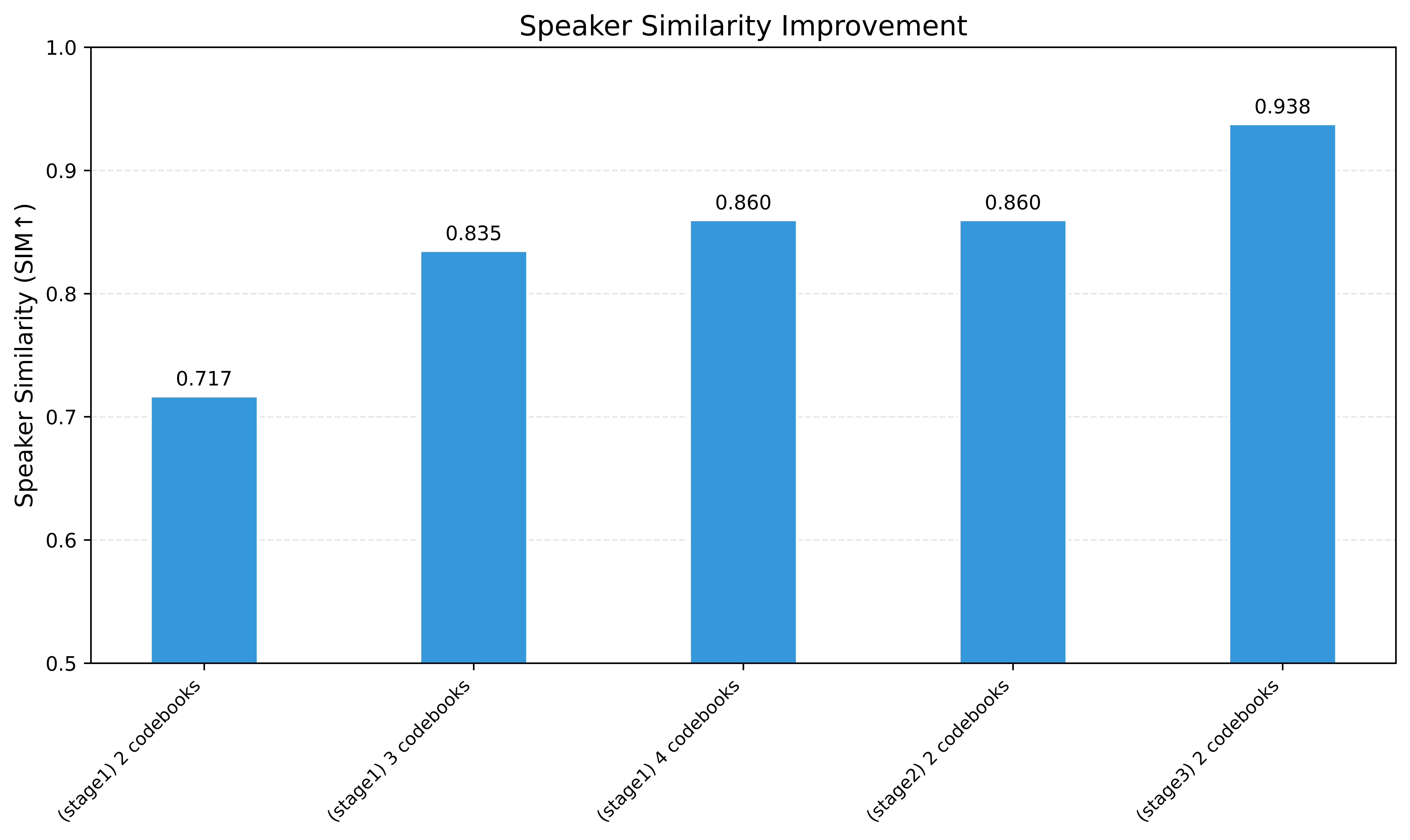}
  \caption{Speaker similarity improvement by Stage 2 and Stage 3}
  \label{fig:speaker_similarity}
\end{figure}

As shown in Figure\,\ref{fig:speaker_similarity}, we observed that the model in Stage 1, when using 2-codebook reconstruction, achieved only moderate speaker similarity. 
This phenomenon can be attributed to two potential reasons: first, the decoder in Stage 1 was required to be compatible with 2, 3, and 4 codebooks simultaneously, thus failing to be optimally tailored for any specific configuration; second, since Stage 1 had no exposure to this particular speaker, the model could only rely on its generalization capabilities for reconstruction. 
During the Stage 2 training phase, the decoder focused on optimizing its 2-codebook reconstruction performance and was also introduced to the target speaker, leading to a rapid increase in speaker similarity from 0.717 to 0.860 (it demonstrates exactly the same speaker similarity as the 4-codebook decoder in Stage 1).
After undergoing Stage 3 fine-tuning with a more focused dataset including the specific speaker, the model further concentrated its reconstruction capability on this target speaker, resulting in a similarity improvement to 0.938.

\paragraph*{Audio Quality}
We used three no-reference objective evaluation metrics to assess the audio quality: 
SIGMOS\,\citep{cutler2024sigmos}\footnote{\url{https://github.com/microsoft/SIG-Challenge}},
NISQA\,\citep{Mittag2021nisqa}\footnote{\url{https://github.com/gabrielmittag/NISQA}},and
UTMOS\,\citep{Saeki2022utmos}\footnote{\url{https://github.com/sarulab-speech/UTMOS22}}. 
SIGMOS and NISQA both evaluate the quality of the speech signal itself from multiple dimensions, including spectral cleanliness, continuity, and noise level et al. 
For the overall score, we adopted the OVRL metric from SIGMOS and the MOS metric from NISQA. 
UTMOS is used to evaluate whether synthesized audio resembles real recorded audio.

\begin{table}[ht]
    \centering
    \begin{threeparttable}
    \caption{Audio quality improvement by Stage 2}
    \label{tab:ob_quality}
    \begin{tabular}{lcccc}
        \toprule
        \textbf{Model} & \textbf{SIGMOS$\uparrow$} & \textbf{UTMOS$\uparrow$} & \textbf{NISQA$\uparrow$} \\
        \midrule
        \textbf{libritts-testset-clean-b} & 3.24 & 4.15 & 4.09 \\
        \textbf{(stage1) 2 codebooks-16k decoder} & 2.94 & 3.85 & 4.20 \\
        \textbf{(stage2) 2 codebooks-24k decoder} & 3.35 & 4.12 & 4.33 \\
        \bottomrule
        \end{tabular}
        
    \end{threeparttable}
\end{table}

As shown in Table\,\ref{tab:ob_quality}, after training a standalone 24kHz decoder on high-quality audio data using 2 codebooks, the quality of the synthesized audio exhibits significant improvements across the three dimensions of SIGMOS, NISQA, and UTMOS. In terms of sound quality, the synthesized audio outperforms that of the Libritts-testset-clean dataset, while in terms of UTMOS, it approaches very closely the quality of real human speech.

These experiments demonstrate that decoupling the training of the Encoder and Decoder enables more flexible control over the Decoder's behavior. 
This allows it to generate audio with high timbre similarity and high quality at extremely low bitrates, thereby adapting more flexibly to the requirements of diverse scenarios.

\section{Conclusion}
In this paper, we introduced LongCat-Audio-Codec, an audio tokenizer and detokenizer solution designed for Speech Large Language Models.
It achieves a low token rate while enabling high-quality, low-complexity streaming audio generation. Our core components mainly include:
\begin{itemize}
    \item \textbf{Efficient Semantic-Acoustic Hierarchical Modeling}: 
    Achieving a balance between comprehension capability and generative capability through hierarchical modeling of the semantic-acoustic tokenizer.
    \item \textbf{Scalable Training with Large Codebooks}:
    Enabling stable training with large codebooks and flexible codebook selection capability via the adaptive codebook grouping strategy and the "train-more-use-less" strategy.
    \item \textbf{Low-Latency, Low-Complexity Generation}: 
    Realizing high-quality speech generation with low latency and low complexity through the elaborately designed decoder and multi-stage training strategy.
\end{itemize}
LongCat-Audio-Codec still has certain limitations in practical application. Specifically, its design is optimized primarily for speech (resulting in insufficient support for music and sound effects), and it currently accepts a maximum audio input length of 30 seconds—requiring pre-slicing for audio exceeding this duration. We will conduct targeted optimizations to address these existing limitations in subsequent versions.

\clearpage  % 强制当前页内容并强制分页
\bibliographystyle{unsrt} % 设置文献排序为引用顺序
\bibliography{longcat_codec}

\begin{thebibliography}{10}

\bibitem{defossez2022encodec}
Alexandre Defossez, Jade Copet, Gabriel Synnaeve, and Yossi Adi.
\newblock High fidelity neural audio compression.
\newblock In {\em Advances in Neural Information Processing Systems}, volume~35, pages 30684--30696, 2022.

\bibitem{kumar2024dac}
Rithesh Kumar, Prem Seetharaman, Alejandro Luebs, Ishaan Kumar, and Kundan Kumar.
\newblock High-fidelity audio compression with improved rvqgan.
\newblock {\em Advances in Neural Information Processing Systems}, 36, 2024.

\bibitem{borsos2023audiolm}
Zal{\'a}n Borsos, Rapha{\"e}l Marinier, Damien Vincent, Eugene Kharitonov, Olivier Pietquin, Matt Sharifi, Dominik Roblek, Olivier Teboul, David Grangier, Marco Tagliasacchi, et~al.
\newblock Audiolm: a language modeling approach to audio generation.
\newblock {\em IEEE/ACM transactions on audio, speech, and language processing}, 31:2523--2533, 2023.

\bibitem{lakhotia2021gslm}
Kushal Lakhotia, Eugene Kharitonov, Wei-Ning Hsu, Yossi Adi, Adam Polyak, Benjamin Bolte, Tu-Anh Nguyen, Jade Copet, Alexei Baevski, Abdelrahman Mohamed, et~al.
\newblock On generative spoken language modeling from raw audio.
\newblock {\em Transactions of the Association for Computational Linguistics}, 9:1336--1354, 2021.

\bibitem{nguyen2024spirit}
Tu~Anh Nguyen, Benjamin Muller, Bokai Yu, Marta~R Costa-Jussa, Maha Elbayad, Sravya Popuri, Paul-Ambroise Duquenne, Robin Algayres, Ruslan Mavlyutov, Itai Gat, et~al.
\newblock Spirit-lm: Interleaved spoken and written language model.
\newblock {\em arXiv preprint arXiv:2402.05755}, 2024.

\bibitem{du2024cosyvoice}
Zhihao Du, Qian Chen, Shiliang Zhang, Kai Hu, Heng Lu, Yexin Yang, Hangrui Hu, Siqi Zheng, Yue Gu, Ziyang Ma, et~al.
\newblock Cosyvoice: A scalable multilingual zero-shot text-to-speech synthesizer based on supervised semantic tokens.
\newblock {\em arXiv preprint arXiv:2407.05407}, 2024.

\bibitem{du2024cosyvoice2}
Zhihao Du, Yuxuan Wang, Qian Chen, Xian Shi, Xiang Lv, Tianyu Zhao, Zhifu Gao, Yexin Yang, Changfeng Gao, Hui Wang, et~al.
\newblock Cosyvoice 2: Scalable streaming speech synthesis with large language models.
\newblock {\em arXiv preprint arXiv:2412.10117}, 2024.

\bibitem{du2025cosyvoice3}
Zhihao Du, Gao Changfeng, Yuxuan Wang, Fan Yu, et~al.
\newblock Cosyvoice 3: Towards in-the-wild speech generation via scaling-up and post-training.
\newblock {\em arXiv preprint arXiv:2505.17589}, 2025.

\bibitem{Julian2024StableCodec}
Julian~D. Parker, Anton Smirnov, Jordi Pons, et~al.
\newblock Scaling transformers for low-bitrate highquality speech coding.
\newblock {\em arXiv preprint arXiv:2411.19842}, 2024.

\bibitem{Ji2024WavTokenizer}
Shengpeng Ji, Ziyue Jiang, Wen Wang, et~al.
\newblock Wavtokenizer: an efficient acoustic discrete codec tokenizer for audio language modeling.
\newblock {\em arXiv preprint arXiv:2408.16532}, 2024.

\bibitem{Ren2024TiCodec}
Zexin Li, Xunying Liu, Shinji Wang, and Ryoichi Yamamoto.
\newblock Fewer-token neural speech codec with time-invariant codes.
\newblock In {\em Proceedings of the 2023 IEEE International Conference on Acoustics, Speech and Signal Processing (ICASSP)}, pages 1--5. IEEE, 2023.

\bibitem{Luo2024LSCodec}
Yiwei Guo, Zhihan Li, Chenpeng Du, et~al.
\newblock Lscodec: Low-bitrate and speaker-decoupled discrete speech codec.
\newblock {\em arXiv preprint arXiv:2410.15764}, 2024.

\bibitem{Hubert2024SNAC}
Hubert Siuzdak, Florian Grötschla, and Luca~A. Lanzendörfer.
\newblock Snac: Multi-scale neural audio codec.
\newblock {\em arXiv preprint arXiv:2410.14411}, 2024.

\bibitem{zhang2023speechtokenizer}
Xin Zhang, Dong Zhang, Shimin Li, Yaqian Zhou, and Xipeng Qiu.
\newblock Speechtokenizer: Unified speech tokenizer for speech large language models.
\newblock {\em arXiv preprint arXiv:2308.16692}, 2023.

\bibitem{defossez2024moshi}
Alexandre D{\'e}fossez, Laurent Mazar{\'e}, Manu Orsini, Am{\'e}lie Royer, Patrick P{\'e}rez, Herv{\'e} J{\'e}gou, Edouard Grave, and Neil Zeghidour.
\newblock Moshi: a speech-text foundation model for real-time dialogue.
\newblock {\em arXiv preprint arXiv:2410.00037}, 2024.

\bibitem{Ye2024codecdoesmatter}
Zhen Ye, Peiwen Sun, Jiahe Lei, Hongzhan Lin, Xu~Tan, Zheqi Dai, Qiuqiang Kong, Jianyi Chen, Jiahao Pan, Qifeng Liu, et~al.
\newblock Codec does matter: Exploring the semantic shortcoming of codec for audio language model.
\newblock {\em arXiv preprint arXiv:2408.17175}, 2024.

\bibitem{Gong2025XYTokenizer}
Yitian Gong, Luozhijie Jin, et~al.
\newblock Xy-tokenizer: Mitigating the semantic-acoustic conflict in low-bitrate speech codecs.
\newblock {\em arXiv preprint arXiv:2506.23325}, 2025.

\bibitem{Liu2024SemantiCodec}
Haohe Liu, Xuenan Xu, Yi~Yuan, et~al.
\newblock Semanticodec: An ultra low bitrate semantic audio codec for general sound.
\newblock {\em arXiv preprint arXiv:2405.00233}, 2024.

\bibitem{Yang2024LLMCodec}
Dongchao Yang, Haohan Guo, Yuanyuan Wang, et~al.
\newblock Uniaudio 1.5: Large language model-driven audio codec is a few-shot audio task learner.
\newblock {\em arXiv preprint arXiv:2406.10056}, 2024.

\bibitem{si2024hertz}
{Standard Intelligence}.
\newblock hertz-dev: the first open-source base model for conversational audio generation.
\newblock \url{https://si.inc/hertz-dev/}, 2024.
\newblock Accessed: 2025-01-16.

\bibitem{Jade2023MusicGen}
Jade Copet, Felix Kreuk, et~al.
\newblock Simple and controllable music generation.
\newblock {\em arXiv preprint arXiv:2306.05284}, 2023.

\bibitem{Yu2023lfq}
Lijun Yu, Jos{\'e} Lezama, Nitesh~B Gundavarapu, Luca Versari, Kihyuk Sohn, David Minnen, Yong Cheng, Vighnesh Birodkar, Agrim Gupta, Xiuye Gu, et~al.
\newblock Language model beats diffusion--tokenizer is key to visual generation.
\newblock {\em arXiv preprint arXiv:2310.05737}, 2023.

\bibitem{Mentzer2023fsq}
Fabian Mentzer, David Minnen, Eirikur Agustsson, and Michael Tschannen.
\newblock Finite scalar quantization: Vq-vae made simple.
\newblock {\em arXiv preprint arXiv:2309.15505}, 2023.

\bibitem{Yang2023hificodec}
Dongchao Yang, Songxiang Liu, Rongjie Huang, Jinchuan Tian, Chao Weng, and Yuexian Zou.
\newblock Hifi-codec: Group-residual vector quantization for high fidelity audio codec.
\newblock {\em arXiv preprint arXiv:2305.02765}, 2023.

\bibitem{chiu2022bestrq}
Chung-Cheng Chiu, James Qin, Yu~Zhang, Jiahui Yu, and Yonghui Wu.
\newblock Self-supervised learning with random-projection quantizer for speech recognition.
\newblock In {\em International Conference on Machine Learning}, pages 3915--3924. PMLR, 2022.

\bibitem{graves2006connectionist}
Alex Graves, Santiago Fern{\'a}ndez, Faustino Gomez, and J{\"u}rgen Schmidhuber.
\newblock Connectionist temporal classification: labelling unsegmented sequence data with recurrent neural networks.
\newblock In {\em Proceedings of the 23rd international conference on Machine learning}, pages 369--376, 2006.

\bibitem{huang2017aishell}
Jinyu Huang, Wei Zhang, Hongji Hao, Di~Xie, Ming Lei, Bin Ma, Lin Zhao, Xunying Liu, Jia Wu, Qing Li, et~al.
\newblock Aishell-1: An open-source mandarin speech corpus and a speech recognition baseline.
\newblock In {\em 2017 IEEE International Conference on Acoustics, Speech and Signal Processing}, pages 5227--5231. IEEE, 2017.

\bibitem{panayotov2015librispeech}
Vassil Panayotov, Guoguo Chen, Daniel Povey, and Sanjeev Khudanpur.
\newblock Librispeech: an asr corpus based on public domain audio books.
\newblock In {\em 2015 IEEE international conference on acoustics, speech and signal processing (ICASSP)}, pages 5206--5210. IEEE, 2015.

\bibitem{nagrani2017voxceleb}
Arsha Nagrani, Joon~Son Chung, and Andrew Zisserman.
\newblock Voxceleb: A large-scale speaker identification dataset.
\newblock In {\em 2017 IEEE International Conference on Acoustics, Speech and Signal Processing}, pages 5329--5333. IEEE, 2017.

\bibitem{gemmeke2017audio}
Jort~F Gemmeke, Daniel P~W Ellis, Dylan Freedman, Aren Jansen, Wade Lawrence, R~Channing Moore, Manoj Plakal, and Marvin Ritter.
\newblock Audio set: An ontology and human-labeled dataset for audio events.
\newblock In {\em 2017 IEEE International Conference on Acoustics, Speech and Signal Processing}, pages 776--780. IEEE, 2017.

\bibitem{busso2008iemocap}
Carlos Busso, Murtaza Bulut, Chi-Chun Lee, Abe Kazemzadeh, Emily Mower, Samuel Kim, Jeannette Chang, Sungbok Lee, and Shrikanth~S Narayanan.
\newblock Iemocap: Interactive emotional dyadic motion capture database.
\newblock In {\em 2008 IEEE International Conference on Acoustics, Speech and Signal Processing}, pages 1069--1072. IEEE, 2008.

\bibitem{Guo2025DiscreteTokens}
Yiwei Guo, Zhihan Li, et~al.
\newblock Recent advances in discrete speech tokens: A review.
\newblock {\em arXiv preprint arXiv:2502.06490}, 2025.

\bibitem{zen2019libritts}
Heiga Zen, Yanqing Dang, Rob Clark, and Colin Raffel.
\newblock Libritts: A corpus derived from librispeech for text-to-speech.
\newblock In {\em Proceedings of the Annual Conference of the International Speech Communication Association, Interspeech}, pages 4986--4990, Graz, Austria, 2019.

\bibitem{Ju2024FACodec}
Zeqian Ju, Yuancheng Wang, Kai Shen, et~al.
\newblock Naturalspeech 3: Zero-shot speech synthesis with factorized codec and diffusion models.
\newblock {\em arXiv preprint arXiv:2403.03100}, 2024.

\bibitem{Wang2023cam++}
Hui Wang, Siqi Zheng, Yafeng Chen, Luyao Cheng, and Qian Chen.
\newblock Cam++: A fast and efficient network for speaker verification using context-aware masking.
\newblock {\em arXiv preprint arXiv:2303.00332}, 2023.

\bibitem{cutler2024sigmos}
Ross Cutler, Ando Saabas, Babak Naderi, Nicolae-C{\u{a}}t{\u{a}}lin Ristea, Sebastian Braun, and Solomiya Branets.
\newblock Icassp 2023 speech signal improvement challenge.
\newblock {\em IEEE Open Journal of Signal Processing}, 2024.

\bibitem{Mittag2021nisqa}
G.~Mittag, B.~Naderi, A.~Chehadi, and S.~Möller.
\newblock Nisqa: A deep cnn-self-attention model for multidimensional speech quality prediction with crowdsourced datasets.
\newblock {\em Proc. Interspeech 2021, 2021}, 2021.

\bibitem{Saeki2022utmos}
Takaaki Saeki, Detai Xin, Wataru Nakata, Tomoki Koriyama, Shinnosuke Takamichi, and Hiroshi Saruwatari.
\newblock Utmos: Utokyo-sarulab system for voicemos challenge 2022.
\newblock {\em arXiv preprint arXiv:2204.02152}, 2022.

\bibitem{kilgour2019frechet}
Kevin Kilgour, Mauricio Zuluaga, and Matthew Sharifi.
\newblock Fr{\'e}chet audio distance: A metric for evaluating music enhancement algorithms.
\newblock In {\em Proceedings of the Annual Conference of the International Speech Communication Association, Interspeech}, pages 2350--2354, Graz, Austria, 2019.

\end{thebibliography}

%%%%%%%%%%%%%%%%%%%%%%%%%%%%%%%%%%%%%%%%%%%%%%%%%%%%%%%%%%%%
\clearpage % 强制分页，确保附录从新页开始
\appendix
\section*{Appendix} % 显示"Appendix"大标题（无编号）
\addcontentsline{toc}{section}{Appendix} % 将附录标题加入目录（可选）

\section{Exploring the Potential of Few Codebooks}
The information carried by 4 codebooks has demonstrated favorable performance in speech-related evaluations. 
We hypothesize that this performance is jointly influenced by the modeling capacity of the encoder and the reconstruction capability of the decoder. To further validate the rationality of the 4 codebooks configuration, we conducted a potential evaluation under the more challenging music scenario.
Specifically, we first trained a 32-codebook acoustic-only music codec, where quantizers were probabilistically deactivated during training to enable the encoder to output continuous latents simultaneously. 
Subsequently, we employed the latents from top 4 codebooks as conditions and a more powerful UNet-based module as Flow Matching decoder. During training , the optimization target set to the ground-truth continuous latents. In the inference phase, we use an Euler solver to perform inference under 30 NFE. After obtaining the predicted Latent, it is fed into the codec decoder for audio reconstruction. FAD\citep{kilgour2019frechet} and multi-resolution STFT distance are evaluated.

\begin{figure}[h]
  \centering
  \includegraphics[width=1.0\textwidth]{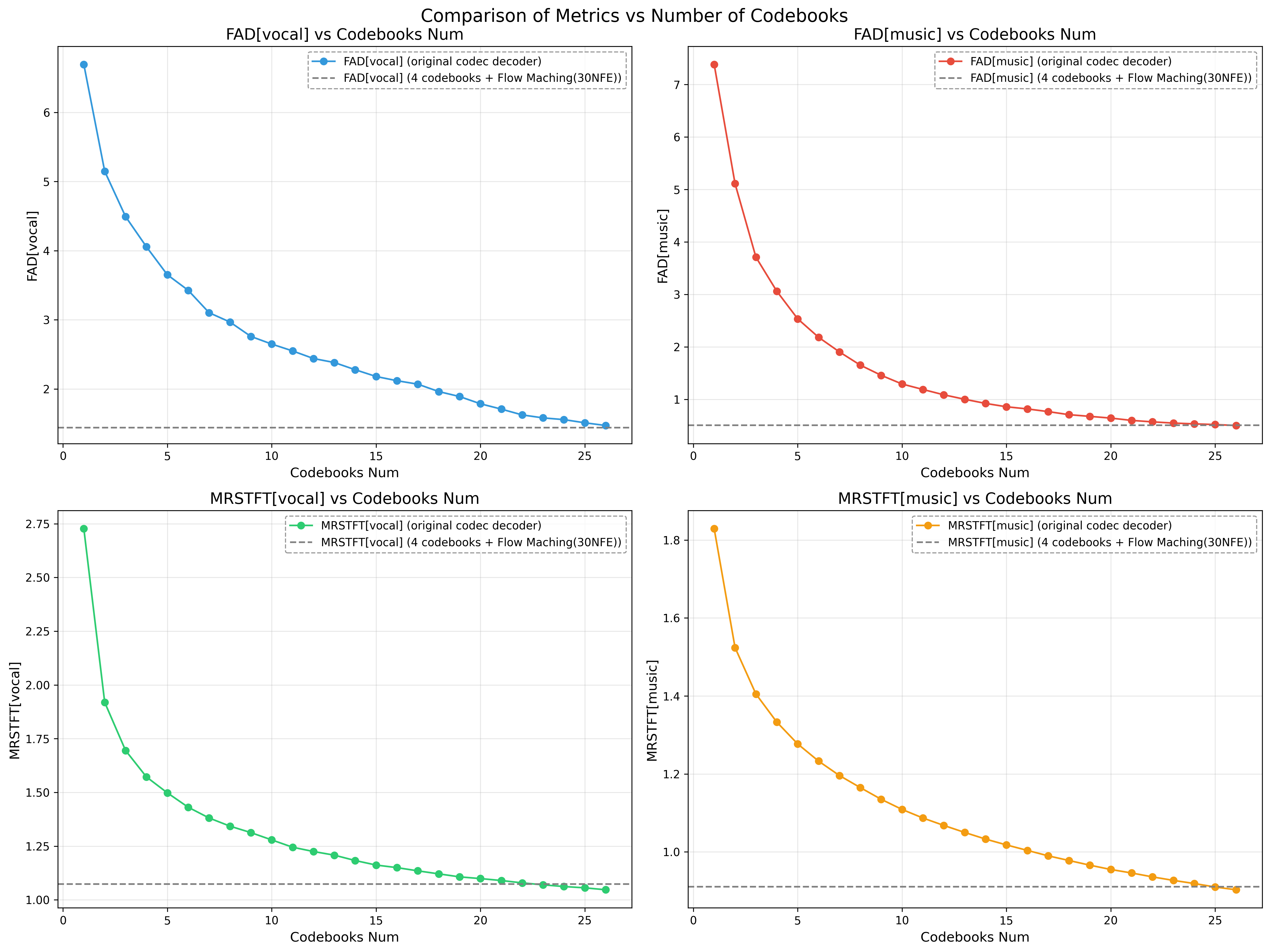}
  \caption{Potential of few codebooks}
  \label{fig:metrics_comparison_plot}
\end{figure}

Figure\,\ref{fig:metrics_comparison_plot} indicate that, with 30 NFE, the information contained in the 4 codebooks can be reconstructed to a quality level equivalent to that of around 22--26 codebooks when paired with a more powerful decoder method (Flow Matching). 
This further confirms that the information capacity of 4 codebooks is sufficiently abundant, and scaling up the decoder can effectively enhance reconstruction performance. 

\end{document}